\documentclass[sn-mathphys-num]{sn-jnl}

\usepackage{amsmath}
\usepackage{braket}
\usepackage{amsfonts}
\usepackage{booktabs}
\usepackage{aas_macros}
\usepackage{caption}
\usepackage{graphicx}
\usepackage{subcaption}
\usepackage{bm}
\usepackage{slashed}
\usepackage{changes}
\usepackage{mathabx}
\usepackage{datatool, filecontents}
\usepackage{hyperref}
\usepackage{xcolor}
\usepackage{graphicx}%
\usepackage{multirow}%
\usepackage{amsmath,amssymb,amsfonts}%
\usepackage{amsthm}%
\usepackage{mathrsfs}%
\usepackage[title]{appendix}%
\usepackage{xcolor}%
\usepackage{textcomp}%
\usepackage{manyfoot}%
\usepackage{booktabs}%
\usepackage{algorithm}%
\usepackage{algorithmicx}%
\usepackage{algpseudocode}%
\usepackage{listings}%

\DeclareMathAlphabet{\mathpzc}{OT1}{pzc}{m}{it}

\def\aap{\textit{Astronomy and Astrophysics}}

\def\be{\begin{equation}}
\def\ee{\end{equation}}
\def\bea{\begin{eqnarray}}
\def\eea{\end{eqnarray}}

\def\L{\mathcal{L}}
\def\Lm{\mathcal{L}_m}

\def\t{\tilde}

\begin{document}

\title[Radiating solutions in Entangled Relativity]{Radiating solutions in Entangled Relativity}

\author*[1,2]{\fnm{Olivier} \sur{Minazzoli}}\email{olivier.minazzoli@oca.eu}
\author[3,1]{\fnm{Maxime} \sur{Wavasseur}}\email{m.wavasseur@protonmail.com}

\affil[1]{\orgdiv{Artemis}, \orgname{Université Côte d’Azur, CNRS, Observatoire Côte d’Azur}, \orgaddress{\city{Nice}, \postcode{BP4229, 06304}, \country{France}}}
\affil[2]{\orgname{Bureau des Affaires Spatiales}, \orgaddress{\street{2 rue du Gabian}, \city{Monaco}, \postcode{98000}, \country{Monaco}}}
\affil[3]{\orgdiv{Departament de Física Quántica i Astrofísica (FQA)}, \orgname{Universitat de Barcelona (UB)}, \orgaddress{\street{Carrer de Martí i Franquès, 1}, \city{Barcelona}, \postcode{08028}, \country{Spain}}}

\abstract{
The Mineur--Vaidya radiating solutions satisfy $\Lm ~\propto~ F^2 = 0 = R$.  
As a consequence, it is not only a solution in General Relativity, but also in Einstein--Maxwell--dilaton theories for all coupling constants.  
The specific case of Entangled Relativity is noteworthy because the additional scalar degree of freedom is defined from the ratio between $R$ and $\Lm$, which is ill-defined in this situation.  
In the present work, we embed the Mineur--Vaidya solution in a magnetic (or electric) field within the framework of Entangled Relativity, and show that the Mineur--Vaidya solution corresponds to the limit where the magnetic (respectively, electric) field vanishes.  
This notably allows us to demonstrate that, as in General Relativity, it is possible to dynamically form naked singularities in Entangled Relativity.  
This conclusion, in fact, applies to any Einstein--Maxwell--dilaton theory, although it does not seem to be widely acknowledged in the literature.
}

\keywords{Radiating solution,naked singularity,beyond general relativity, Einstein--Maxwell--dilaton theories}



\maketitle
\section{Introduction}

Entangled Relativity is a simple non-linear reformulation of General Relativity \cite{ludwig:2015pl,minazzoli:2025ar}.  
Its strength lies in being more parsimonious in terms of fundamental constants than General Relativity---thanks to its non-linear coupling (see Sec.~\ref{sec:ER})---while recovering General Relativity without a cosmological constant as a limit in the near-vacuum regime and for all matter fields that satisfy $\Lm = T$ on-shell \cite{minazzoli:2021ej,arruga:2021pr,arruga:2021ep,minazzoli:2021cq,minazzoli:2025ep,minazzoli:2025ej,minazzoli:2025ar}.  
Since this condition holds for a universe composed of dust and radiation, it follows that the phenomenology of Entangled Relativity is very close to that of General Relativity in the observable universe.  
Moreover, it has recently been shown that Entangled Relativity is the only $f(R,\Lm)$ theory, apart from General Relativity itself, that possesses these characteristics \cite{minazzoli:2025ar}.  
Therefore, Entangled Relativity represents an interesting alternative to General Relativity. \\

The non-linearity of the coupling between matter and curvature in the action induces the presence of an additional scalar degree of freedom in the field equations—as is usual in $f(R)$ theories \cite{capozziello:2015sc}.  
However, due to the specific non-linearity between $R$ and $\Lm$ in the field equations inherited from the `action'\footnote{Let us note that we use the name `\textit{action}' for historical reasons, but the spacetime integral over the Lagrangian density in Eq. (\ref{eq:SER}) actually has the dimension of an energy squared \cite{minazzoli:2022ar,chehab:2025hl}, because the Lagrangian density itself has the dimension of an energy squared over a length to the fourth---see Eq. (\ref{eq:SER}).}---see Eqs. (\ref{eq:SER}-\ref{eq:vartheta})---it was argued in \cite{minazzoli:2021ej} that the scalar degree of freedom might become negative in some situations.
Hence, it was speculatively argued in \cite{minazzoli:2021ej} that Entangled Relativity might be able to resolve the singularity problem of classical relativistic theories, through repulsive gravity. \\

In the present paper, we find explicit solutions that form not only singularities, but also (locally or globally) naked singularities in Entangled Relativity.  
The demonstration is straightforward: it starts from the Melvin--Vaidya radiating solutions of General Relativity---known to produce all the singularities mentioned above---and adapts them to Entangled Relativity by embedding them in either a magnetic or an electric field. The resulting solutions are straightforwardly shown to be able to form hidden or naked singularities as well.

\section{Vaidya are solutions of all Einstein--Maxwell--dilaton theories}

In convenient units, the action of Einstein--Maxwell--dilaton theories can be written as follows \cite{garfinkle:1991pr,holzhey:1992nb}:
\bea \label{eq:actionEMd}
S=\int d^{4} x \sqrt{-\t g}\left[\t R-2\t g^{\mu \nu}\partial_\mu \varphi \partial_\nu \varphi -  e^{- 2\alpha\varphi}  \t F^2 \right],
\eea
where $\alpha$ takes a specific value for each specific theory. For instance, it takes the value $\alpha=1$ for bosonic strings \cite{holzhey:1992nb}, $\alpha = \sqrt{3}$ for a 5D Kaluza-Klein theory with one dimension compactified on a circle \cite{holzhey:1992nb} and more generally $\alpha = \sqrt{(2+N)/N}$ for a (4$+N$)D Kaluza-Klein theory with a compactified N-dimensional torus \cite{gibbons:1988nb}, whereas $\alpha=1/(2\sqrt{3})$ for Entangled Relativity \cite{minazzoli:2021ej}.

\subsection{Mineur--Vaidya radiating solutions}

The so-called Vaidya radiating solution \cite{vaidya:1953na} has been first derived by Mineur \cite{mineur:1933an}, albeit, in a different coordinate system \cite{Gourgoulhon:BHnotes}. For ingoing and outgoing rays, it reads as follows 
\be
d\t s^2 = - \left(1-\frac{r_s(\xi)}{r} \right) d\xi^2 + 2 \epsilon d\xi dr + r^2 d\Omega^{2},
\ee
where $d\Omega^{2}=(d\theta^2 + \sin^2 \theta d\psi^2)$, the Eddington-Finkelstein coordinates  $\xi := r+\epsilon t$ and $r_s(\xi)$ the Schwarzchild radius which depends on $\xi$. The ingoing and outgoing solutions correspond to the cases $\epsilon=1$ and $\epsilon=-1$ respectively. They are solutions to the equation of General Relativity without a cosmological constant that is sourced by a spherical symmetric electromagnetic radiation within the geometric optics approximation:\\
\be
\t T^{\mu \nu}(\xi) =  \epsilon\frac{\dot r_s(\xi)}{r^2} k^\mu k^\nu,
\label{eq:Mineur_vaidya_ener_imp}
\ee
with $k_\alpha:= - \partial_\alpha \xi$ and where $\dot r_s(\xi)$ stands for the derivative of the function $r_s(\xi)$. The Mineur--Vaidya solution is such that $\t R=0$ everywhere, since the metric field is sourced by a conformally invariant (null) field $\t T =0$, since $k_\sigma k^\sigma =0$.

\subsection{Solution in Einstein--Maxwell--dilaton theories}

From the geometric optics approximation, one deduces that $\t F^2 = 0 = \t T$,\footnote{From $\mathbf{A} = \mathfrak{R} \{ i a e^{i \omega} \mathbf{f}\}$, the Faraday tensor reads $ \mathbf{F} = \mathfrak{R} \{ i a e^{i \omega} \mathbf{k} \wedge \mathbf{f} \}$, with $\mathbf{k}:=\mathbf{\nabla} \omega$ and $\mathbf{f}$ the wave and polarization vectors respectively, which are such that $\mathbf{k} \cdot \mathbf{k}=0$ (null vector) and $\mathbf{k} \cdot \mathbf{f}=0$ (polarization orthogonal to the rays). $\t F^2 = 0 = \t T$ directly follows from that, in the geometric optics assumption \cite{MTW}.} such that the scalar field equation that derives from the Einstein--Maxwell--dilaton action in Eq. (\ref{eq:actionEMd}) is not sourced. As a consequence, the Mineur--Vaidya radiating solutions of General Relativity are also solutions of Einstein--Maxwell--dilaton theories with a constant dilaton field. We shall come back to this in Sec. \ref{sec:singEMd}.\\

Entangled Relativity is a specific case of Einstein-dilaton theories, which can be rewritten in the Einstein frame in the form of Eq. (\ref{eq:actionEMd}) with the coupling constant $\alpha=1/(2\sqrt{3})$ \cite{minazzoli:2021ej,arruga:2021ep}, provided that $(\Lm,R) \neq 0$. Indeed, the field equations of Entangled Relativity in the conformal fame of its definition---the \textit{entangled frame} \cite{minazzoli:2025ej}---depend on a scalar-field degree of freedom defined from the ratio between $R$ and $\Lm$---see Sec. \ref{sec:ER}.\\

As a consequence, the Mineur–Vaidya solutions cannot, strictly speaking, be solutions of Entangled Relativity.\footnote{The same issue applies for the $\lambda_5$ wormhole solution in \cite{bixano:2025pr,bixano:2025ar,bixano:2025as}, which is such that $\t \L_m =0$.}
In what follows, we first embed the Mineur–Vaidya solutions in a magnetic or electric field within the framework of Einstein–Maxwell–dilaton theories.  
This provides solutions that are such that $\t \L_m \neq 0 \neq \t R$.  
We then verify that they are also solutions of Entangled Relativity after an inverse conformal transformation, and show that the Mineur–Vaidya solutions correspond to the limit of these new solutions when the magnetic (respectively, electric) field tends to zero.

\section{Mineur--Vaidya--Melvin solutions in Einstein--Maxwell--dilaton theories}
\label{sec:MVMEMd}

Using the procedure of \cite{dowker:1994pr}, let us embed the Mineur--Vaidya radiating solutions in a magnetic or electric field in Einstein--Maxwell--dilaton theories. The metric field reads
\be
d\t s^2 = \left[- \left(1-\frac{r_s(\xi)}{r} \right) d\xi^2 + 2 \epsilon d\xi dr + r^2 d\theta^2\right] \Lambda^{\frac{2}{\alpha^2 +1}} + \frac{r^2\sin^2 \theta}{\Lambda^{\frac{2}{\alpha^2+1}}} d\psi^2.
\ee

\subsection{Magnetic case}
In the magnetic case, one has
\be \label{eq:lamMag}
\Lambda = 1+ \frac{1+\alpha^2}{4} B^2 r^2 \sin^2 \theta,
\ee
while the additional Maxwell field reads
\be \label{eq:EMdB}
A = -\frac{2}{(1+\alpha^2)B \Lambda} d\psi.
\ee
corresponding to a magnetic field pointing along the z-direction that reads
\be
\mathfrak{B}^{\mu} =  \frac{16B r \sin\theta }{\left(B^{2} r^{2} \sin^{2}\theta \left(\alpha^{2} +1\right)+ 4\right)^{2}}\left[r \cos\theta dr-\sin\theta d\theta\right],
\ee
It is solution to the Einstein--Maxwell--dilaton that is sourced by the dilaton field
\be
\varphi = -\frac{\alpha}{1+\alpha^2} \ln \Lambda,
\ee
and by both the magnetic field Eq. (\ref{eq:EMdB}) and an axially symmetric electromagnetic radiation within the geometric optics approximation:
\be
\t T^{\mu \nu}(\xi) = \epsilon \frac{\dot r_s(\xi)}{r^2}~\Psi~ k^\mu k^\nu,
\ee
with $\Psi := \Lambda^{-\frac{2\alpha^2}{1+\alpha^2}}$. The outgoing solution can be deduced from the ingoing solution straightforwardly.\\

The notebook that verifies this solution is freely accessible at: \cite{notebook_EMd_mag}.
\subsection{Electric case}

The scalar and electromagnetic field solutions in the case of a uniform electric field can be obtained through the following transformation \cite{garfinkle:1991pr,holzhey:1992nb}

\begin{subequations}\label{eq:transfoEF}
\bea
&& \t F_{\mu \nu}  \longrightarrow \t F^{e}_{\mu \nu} = \frac{1}{2} e^{-2\alpha \varphi}\epsilon_{\mu \nu \kappa \lambda} \t F^{\kappa \lambda},\\
&& \varphi \longrightarrow \varphi^{e} = - \varphi,
\eea
\end{subequations}
where $\epsilon_{\mu \nu \kappa \lambda}$ is the Levi-Cività tensor. \\

The electromagnetic four potential is now given by:
\be
A(\xi) = -B(r -r_{s}(\xi)) \cos\theta  \mathrm{d} t + \epsilon B r_{s}(\xi)\cos\theta \mathrm{d} r.
\ee
Note that this potential is now independent of $\alpha$ and corresponds to an electric field pointing along the z-direction that reads
\be
\label{eq:elec_emd}
\mathfrak{E}^{\mu}(\xi) =  B \cos\theta dr - B\left(r - r_{s}(\xi)\right) \sin\theta d\theta.
\ee
Again, one can verify that in the $B\rightarrow0$ limit, we have $A\rightarrow0$ and the line element converges to the Mineur-Vaidya solution.\\

This solution has been verified in the following notebook: \cite{notebook_EMd_elec}.

\section{The special case of Entangled Relativity in the Entangled frame}
\label{sec:ER}

In convenient units, the action of Entangled Relativity in the frame of its formulation---which we shall hereafter name the ``\textit{entangled frame}'' \cite{minazzoli:2025ej}---reads as follows \cite{ludwig:2015pl}
\be \label{eq:SER}
S = -\frac{1}{2}\int d^4 x \sqrt{-g} \frac{\Lm^2}{R}.
\ee
The metric-field equation that follows from the extremization of the action is \cite{ludwig:2015pl}
\be \label{eq:metric}
G_{\mu \nu} = \frac{T_{\mu \nu}}{\vartheta} + \vartheta^{-2} \left[\nabla_\mu \nabla_\nu - g_{\mu \nu} \Box \right] \vartheta^{2},
\ee
with $\vartheta$ a scalar defined upon the ratio between the scalars $\Lm$ and $R$
\bea
\vartheta := - \frac{\Lm}{R}\label{eq:vartheta}.
\eea
The action in Eq. (\ref{eq:SER}) can be rewritten in the Einstein frame as Eq. (\ref{eq:actionEMd}) with $\alpha = 1/(2\sqrt{3}$), provided that $\Lm \neq \emptyset$ \cite{minazzoli:2021ej,minazzoli:2025ej}. Therefore, one just has to do an inverse conformal transformation $g_{\alpha \beta} = e^{4\alpha \varphi} \t g_{\alpha \beta}$ with $\alpha=1/(2\sqrt{3})$ in order to deduce the solution of the theory defined in Eq. (\ref{eq:SER}) from the solutions given in Sec. \ref{sec:MVMEMd}. Let us stress that, because of Eqs. (\ref{eq:metric}-\ref{eq:vartheta}), the solutions must be such that $(\Lm,R) \neq 0$ to be well-defined.

\subsection{Magnetic case}
The ingoing and outgoing metric field solutions read
\be
d s^2 = \left[- \left(1-\frac{r_s(\xi)}{r} \right) d\xi^2 + 2 \epsilon d\xi dr + r^2 d\theta^2\right] \Lambda^{\frac{20}{13}} + \frac{r^2\sin^2 \theta}{\Lambda^{\frac{28}{13}}} d\psi^2.
\ee
with
\be
\Lambda = 1+ \frac{13}{48} B^2 r^2 \sin^2 \theta,\textrm{ and, }A = -\frac{24}{13 B \Lambda} d\psi.
\ee
It is solution to the Entangled Relativity field equations with both the magnetic field defined in Eq. (\ref{eq:EMdB}) and an axial symmetric electromagnetic radiation within the geometrical optics approximation:
\be\label{eq:tmunuERMag}
T^{\mu \nu}(\xi) = \frac{\dot r_s(\xi)}{r^2}~\vartheta~ k^\mu k^\nu,
\ee
The on-shell values of the Ricci scalar and matter Lagrangian are
\be
R(\xi) = - B^2~ \frac{r_s(\xi) \sin^2 \theta  - r^2}{\Lambda^{\frac{46}{13}}},
\ee
and
\be
\Lm(\xi) = B^2~ \frac{r_s(\xi) \sin^2 \theta  - r^2}{\Lambda^{\frac{44}{13}}},
\ee
such that
\be
\vartheta := -\frac{\Lm}{R} = \Lambda^{\frac{2}{13}},
\ee
where $\Lambda$ is given by Eq. (\ref{eq:lamMag}). Let us stress that the stress-energy tensor in Eq. (\ref{eq:tmunuERMag}) acquires an axial symmetry with respect to the usual spherical Mineur-Vaidya solution given in Eq. (\ref{eq:Mineur_vaidya_ener_imp}), from the presence of the magnetic field.\\

The verification of this solution can be found in the following notebook: \cite{notebook_ER_mag}.

\subsection{Electric case}
The electric version of the solution can be obtained through the following transformation \cite{minazzoli:2024ax}
\begin{subequations}\label{eq:transfo}
\bea
&&F_{\mu \nu}  \longrightarrow F^{e}_{\mu \nu} =-\frac{1}{2} \frac{\Lm}{R} ~ \epsilon_{\mu \nu \kappa \lambda} F^{\kappa \lambda},\\
&&g_{\mu \nu} \longrightarrow g^e_{\mu \nu} = \left(\frac{\Lm}{R}\right)^4 g_{\mu \nu},
\eea
\end{subequations}\label{eq:litdiffm}
where $\epsilon_{\mu \nu \kappa \lambda}$ is the Levi-Cività tensor, such that \cite{wavasseur:2025gg}
\be \label{eq:transfoERmag2elec}
\vartheta \longrightarrow \vartheta^e = \frac{1}{\vartheta}.
\ee
Let us stress that, unlike in the Einstein frame, the line element in the electric case differ from the magnetic one.
 Hence, in the case of a uniform electric field, the solution for the metric is: 
\be
d s^2 = \left[- \left(1-\frac{r_s(\xi)}{r} \right) d\xi^2 + 2 \epsilon d\xi dr + r^2 d\theta^2\right] \Lambda^{\frac{28}{13}} + \frac{r^2\sin^2 \theta}{\Lambda^{\frac{20}{13}}} d\psi^2.
\ee
with the same four potential than in Einstein-Maxwell-dilaton theories, corresponding to an electric field pointing along the z-direction. The ricci and on-shell matter Lagrangian scalars transform as follows with respect to the magnetic case
\bea
R &\longrightarrow& R^e=-R/\vartheta^4, \\
\Lm &\longrightarrow& \Lm^e=-\Lm/\vartheta^6.
\eea
Their ratio therefore retains the same sign in both cases.\\

This solution has been verified in the following notebook: \cite{notebook_ER_elec}.

\section{Discussion on the formation of singularities}
\label{sec:singEMd}

Mineur–Vaidya collapses are known not only to generate hidden singularities, but also locally or even globally naked singularities \cite{steinmuller:1975pl,griffiths:2009bk,Gourgoulhon:BHnotes}.  
Moreover, their resulting Cauchy horizons have been shown to be stable against non-spherical perturbations \cite{nolan:2005pr,nolan:2007cq}.  
Since they are also solutions of Einstein–Maxwell–dilaton theories for all values of $\alpha$, this demonstrates that all Einstein–Maxwell–dilaton theories can form naked singularities, just as General Relativity does.  
This result trivially extends the conclusion of \cite{aniceto:2016jh}, where it was argued that only the Einstein–Maxwell–dilaton theory with coupling $\alpha = 1$ could lead to naked singularities. \\

The case of the Mineur–Vaidya–Melvin collapse is no different, despite its lack of spherical symmetry.  
Indeed, one can study the singularity formation along the axis of symmetry ($\theta = 0$) and apply all derivations from the spherical Mineur–Vaidya case \cite{steinmuller:1975pl,griffiths:2009bk,Gourgoulhon:BHnotes} to demonstrate the formation of singularities.  
In particular, this shows that Entangled Relativity can also form singularities---and even locally or globally naked ones---although it has been argued that it might not have been the case in \cite{minazzoli:2021ej}. \\

Of course, as is usual with solutions that violate the Cosmic Censorship Conjecture, one may question the physical plausibility of the assumed matter content \cite{wald:1984bk}, here an electromagnetic radiation in the geometric optic limit.

%
\bmhead{Acknowledgements}
The authors gratefully acknowledge Eric Gourgoulhon for the excellent computational notebooks and the exceptional quality of his freely available lecture notes on \textit{Geometry and Physics of Black Holes} \cite{Gourgoulhon:BHnotes}. The codes that verify the field equations in the present paper use SageManifolds \cite{gourgoulhon:2015jc} and are available on GitHub at: \url{https://github.com/mWavasseur/ER/tree/main/Art.IV\%20Radiating\%20solutions\%20in\%20ER}.\\

\textbf{Data Availability Statement}: No data associated in the manuscript.

\bibliography{ER_Melvin_Vaidya}

\end{document}